# Negative Capacitance Ion-Sensitive Field-Effect Transistors with improved current sensitivity

Francesco Bellando*, Ali Saeidi, and Adrian M. Ionescu

[1]Ecole Polytechnique Federale de Lausanne, Lausanne, Switzerland

**Ion-Sensitive Field-Effect Transistors (ISFETs) form a wide-spread technology for sensing, thanks to their label-free detection and intrinsic CMOS compatibility. Their current sensitivity, $\Delta I_D/I_D$, for a given $\Delta pH$, however, is limited by the thermionic limit for the Subthreshold Slope (SS) of Metal-Oxide-Semiconductor Field-Effect Transistors (MOSFET) and by the Nernst limit.**
Obtaining ISFETs with a steep slope transfer characteristics is extremely challenging. In this paper we combine the merits of traditional ISFETs with the performance boosts offered by the insertion of a Negative Capacitor in series with the Gate contact.
The effect of the Negative Capacitance (NC) is beneficial from two points of view: (i) first, it lowers the SS, which provides increased current sensitivity in the weak inversion regime, and (ii) it provides a significant overdrive of the sensor, which reduces its power consumption.
In the proposed tests with NC PZT capacitors, we demonstrate experimentally a reduction of the SS by 44%, combined with a current efficiency improvement of more than two times. As a consequence of the steeper SS, the current sensitivity to pH is improved by 78%. False positive results, in which an improvement in the SS is observed because of secondary effects, are screened out measuring the differential gain of the Ferroelectric Capacitor (FE-Cap), which is bigger than 1 only when the true NC effect arises. We also performed tests with a non-capacitively matched Ferroelectric ISFET and observed no improvement in the SS, coherently with the claims that only the true Negative Capacitance effect can provide the differential gain needed for the performance improvements. In the last part of the paper, we investigate a new sensing method for pH in NC ISFET, based on the differential gain given by the NC: the peak of this characteristics is in fact observed to shift monotonically with the pH of the Liquid Under Test (LUT), the same way as the threshold voltage ($V_{th}$) of the ISFETs does.
Overall, this paper is the first experimental demonstration of performance enhancements of ISFETs through exploitation of the NC effect.

The possibility of using FETs for sensing has been known since the paper of Piet Bergveld in1970[1]. After that first break-through, the advantages given by their low fabrication cost, low power consumption, and intrinsic CMOS compatibility have been exploited in many studies[2–12].
In its conventional configuration, an ISFET has a structure almost identical to that of a MOSFET, the only difference being that the Gate metal contact is removed and substituted with a liquid Gate, composed by the LUT and a Reference Electrode (RE) which applies the bias. This modification allows the ions present in the solution to influence the transconductance of the channel thanks to the difference between their concentration at the interface with the Gate (buffered by the amphoteric sites of the sensing layer) and in the bulk of the LUT[13]. The equation relating the Drain current ($I_D$) of the device to the Gate bias is therefore modified to keep into account this contribution.
Equation (1) shows the new relation in weak inversion condition, which gives the strongest relative current dependence on the Gate bias.

$$I_D = I_{D0} exp\left[\frac{q}{nkT}\left(V_{GS} - \gamma - 2.3\alpha\frac{kT}{q}pH\right)\right]. \tag{1}$$

In this equation, q is the elementary charge, n is the ideality factor of the device (n ≤ 1 for conventional FETs and ISFETs), kT is the thermal energy, $V_{GS}$ is the Gate-Source Voltage, γ is a constant taking into account the RE potential and α is a parameter (α ≤ 1) which denotes the ion-buffering capabilities of the gate material.
Achieving label-free measurements with extremely scalable sensors and low power consumption has proven especially intriguing for wearable devices and POC analysis. Recently, Shogo Nakata et al. demonstrated on-skin testing of a flexible ISFET-based wearable device for continuous monitoring of pH[14]. Matthew Douthwaite showed a similar device, but with a focus on energy harvesting techniques based on thermoelectricity[15]. A wearable device with an integrated miniaturized Quasi-RE (QRE) and functionalized ISFET gates, able to simultaneously monitor pH and the concentration of Sodium and Potassium ions, has also been presented[16] and many other projects have been carried out on similar topics[17–19]. One of the challenges that must be faced, when moving from an expensive and time-consuming laboratory analysis to a fast, cheap and potentially real-time POC one, is the loss of precision due to the limited complexity of sensor and read-out systems, as well as to the added noise coming from a non-controlled environment. In biological analysis, even small relative variations in the concentration of a biomarker can be relevant[20] which brings a continuous need of systems with improved sensitivity. Recently, negative capacitance of ferroelectric materials has been proposed as a mean to step up the gate voltage and hence, improve the SS of field-effect transistors[21, 22]. It is well-established that the employ of a FE-Cap in-series with the Gate of a field-effect transistor could offer an internal voltage amplification, which has been proven to reduce the SS below the thermal limit of conventional MOSFETs[23]. The NC region of ferroelectrics is unstable, showing hysteretic jumps in the polarization. However, this region can be stabilized if the FE-Cap is placed in-series with a positive capacitor of proper value[24]. In this study, we investigate the potentials of the application of the true Negative Capacitance effect to improve the pH response of conventional ISFETs. A Pb(Zr,Ti)O$_3$ (PZT) Fe-Cap, as the NC booster, is placed between the Gate bias source and RE. In a well-designed structure, the NC of PZT is fully stabilized by the series capacitance of the ISFET, performing an almost hysteresis-free behavior. As a result of the true NC effect, the SS of the ISFET is reduced by 44% and the pH sensing is consequently improved by 78%.

**Principle** The overall sensitivity of an ISFET relies on two transductions: (i) the first is relating the ion concentration in the solution to the $V_{th}$ shift and is determined by the buffer capacitance of the material used for sensing, and, (ii) the second is the one relating the $V_{th}$ shift to the $I_D$ level variation, which depends on the efficiency of the gate coupling of the device. The insertion of an NC booster in series with the sensing gate (Figure 1a) is meant to improve the sensing ability of ISFETs acting on this second transduction by providing voltage amplification. This results in steeper ISFET characteristics, meaning that the same amount of pH variation leads to a stronger change of $I_D$ compared to the conventional device (Figure 1b). In this embodiment, the FE-Cap is used to modify the bias applied to the entire LUT, potentially allowing using a single NC element for a full set of ISFETs individually functionalized, as the one reported in reference[16]. Furthermore, this configuration allows placing the FE-Cap, often composed by polluting materials, safely separated from the LUT.

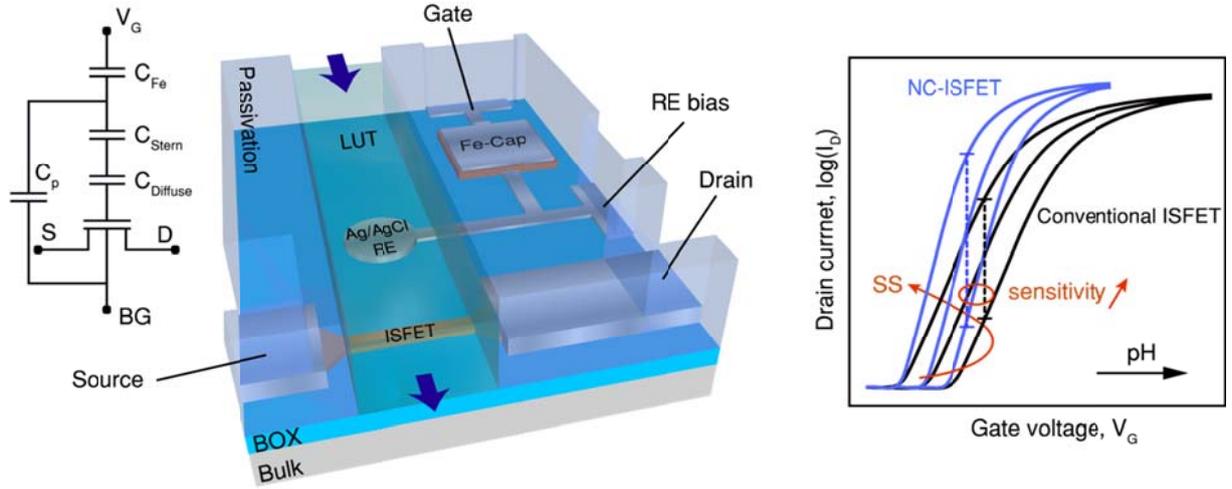

Figure 1 Negative Capacitance ISFET. a) Envisioned embodiment of the integrated NC-ISFET with equivalent circuit for capacitances: the Fe-Cap infers its gain to the Gate bias, which is then applied to the LUT via the integrated RE. b) Expected effect of the NC configuration: though the shifts between the transfer characteristics are not changed, Δ$I_D$/Δ$pH$ is increased by the steeper SS.

Negative capacitance in ferroelectric materials arises from the imperfect screening of the spontaneous polarization, which is intrinsic in semiconductor-ferroelectric and metal-ferroelectric structures. The physical separation of ferroelectric bound charges provides a depolarization field inside the ferroelectric, which destabilizes the polarization. Hence, intentionally destabilizing the polarization causes an effective NC[25]. The concept of NC can be clearly understood by considering the free energy landscape of ferroelectrics, which is traditionally modeled with a double-well energy function (Figure 2a, black curve). The free energy of ferroelectric materials is calculated by the Landau equation (2)

$$U_{FE} = \alpha P_{FE}^2 + \beta P_{FE}^4 + \gamma P_{FE}^6 + E_{EXT} P_{FE}, \qquad (2)$$

Where $U_{FE}$ is the free energy of the ferroelectric, P is the polarization, $E_{EXT}$ is the externally applied electric field, and α, β and γ are material dependent parameters. Generally, the ferroelectric resides in one of the wells in equilibrium and provides spontaneous polarization. The curvature of the U-P characteristic of the ferroelectric is positive around the wells and so is its capacitance, defined in equation (3).

$$C_{FE} = \left[\frac{d^2 U_{FE}}{dP_{FE}^2}\right]^{-1}. \qquad (3)$$

Nevertheless, this curvature is effectively negative when it is moving from one stable polarization state to the other one, which was previously called as the polarization destabilization. As a consequence, the Landau equation foresees an interval of $E_{EXT}$ values in which $P_{FE}$ is expected to decrease when $E_{EXT}$ is increased (center part of the red curve in figure 2b), meaning that the differential capacitance of the ferroelectric is effectively negative.

As stated, the NC has been proven elusive for ferroelectrics in isolation, exhibiting hysteretic jumps in the polarization (Figure 2b, black curve). However, a FE-Cap can be stabilized and provide voltage amplification if it is placed in-series with a positive capacitor of proper value[26], so that the overall energy minimum of the two-capacitor system lies in the point of zero polarization, corresponding to the NC region of the ferroelectric (Figure 2a, red curve). The FE-Cap can operate as a step-up voltage transformer in the NC regime (Figure 2c). Specifically, the FE-Cap brings an abrupt increase in the differential charge of the internal node, between the FE-Cap itself and the stabilizing series capacitor (Figure 2d). A FE-Cap demonstrating a true Negative Capacitance effect in series with a conventional ISFETs can possibly improve the SS and, hence, the sensitivity of the structure.

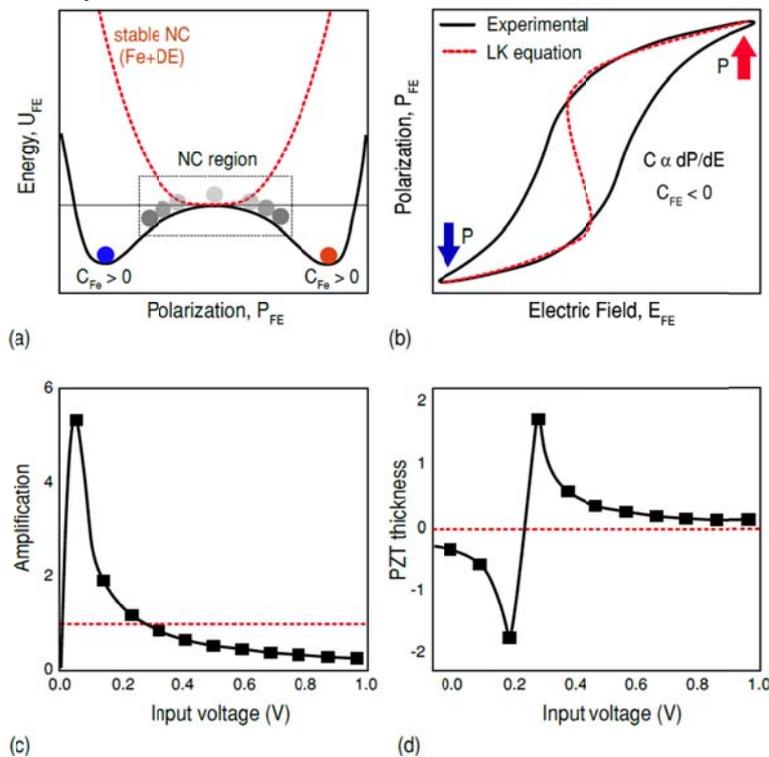

Figure 2 Negative Capacitance of ferroelectrics. a) Energy landscape of isolated and capacitively coupled Fe-Cap: when capacitive matching is achieved the most favorable configuration is that in the NC region. b) Comparison between the shape of the real polarization-electric field characteristics and the one foreseen by the Landau-Khalatnikov equation for an isolated Fe-Cap, showing how the expected Negative Capacitance is not observed and substituted by a hysteretic jump c) Amplification factor evaluated with our model for liquid Gate capacitance. d) Capacitance-voltage characteristics of the Fe-Capacitor, showing the jump due to polarization switching.

However, providing a fully-matched design that ensures a positive total energy for the system in the whole range of operation[23] is extremely challenging considering the fact that both ferroelectric and ISFET are voltage-dependent capacitors. Additionally, the complexity deriving from the employ of a liquid Gate should be considered. Therefore, a detailed study of the physics of the complete system is needed. We have employed the presented capacitance model of Figure

1a) to theoretically investigate the impact of a negative capacitance booster on the ISFET operation. Our model was based on the study reported by Velikonja et al.[27], in which the capacitance of the liquid in contact with an electrode is described as the series of the one of the Stern Layer (i.e. the thin layer composed by the molecules which bind to the electrode) and the one of the Diffuse Layer (i.e. the part of the liquid in which molecules are free to diffuse). The capacitance seen by the Fe-Cap is therefore given by the series of the Diffuse Layer capacitance, the Stern Layer capacitance, the Gate oxide capacitance and the Silicon depletion layer capacitance, all in parallel with the parasitic capacitance given by the measurement setup. The Stern Layer and Diffuse Layer capacitances both depend non-linearly on the electric field falling on the Stern Layer, which in turns depends on both capacitances due to the capacitive voltage divider effect. An analytic solution is therefore not possible and the capacitance of the Double Layer will have to be evaluated self-consistently for each combination of Gate bias value and LUT pH. The FE-Cap behavior is modeled using the Landau-Khalatnikov theory of ferroelectrics[28]. We postpone the detailed mathematical analysis of the total capacitance seen by the Fe-Cap to the Supplementary Information section and report here the numerical solutions, obtained as a function of the applied bias, for the NC gain $\delta V_{int}/\delta V_{out}$ (Figure 2c). The pH of the solution is set to 4.

**Experiments** NC matching has been attempted with a Fully-Depleted Ultra-Thin Body Silicon On Insultator (FD UTB SOI) ISFET which we fabricated in the CMi cleanroom: this ISFET, while showing a poor metal gate SS of 118 mV/dec, is interesting from a point of view of capacitive matching thanks to its fully depleted nature, which makes its equivalent capacitance independent from the external bias. An Ultra-Thin Body Silicon On Insulator (UTB SOI) ISFET has been fabricated to test the foreseen performance improvements. The fabrication process of the device is depicted in Figure 3a). The starting substrate is an ultra-thin (70 nm thick) SOI resting on a BOx with a thickness of only 20 nm, to improve the control via the backgate bias ($V_{BG}$). The SOI layer is further thinned to 30 nm with an oxidation-etch cycle to achieve good electrostatic control. Photolithography has been used to define the phosphorus implantation areas for S and D and, after thermal annealing, has been employed again to define the physical shape of the devices before HBr plasma etching. A thin layer (5 nm thick, according to the TEM measurement) of SiO2 has been grown before the Atomic Layer Deposition (ALD) of the gate dielectric (HfO2) to improve the interface quality, reducing charge trapping and hysteresis. Ti-Pt gates have been deposited by lift-off on the FET channels.
Photolithography and Ion Beam Etching (IBE) have been employed to open holes in the dielectric layer over the S and D areas, to allow electrical connection of the devices with sputtered Pt metal lines. Excellent ohmic contacts have been achieved on 30 nm Si film. A passivation SU-8 layer has been spin-coated on the devices to insulate the metal lines from the LUT. The openings for the sensing gates have been photolithographically defined (Figure 3d).
For the fabrication of the Fe-Cap, a 46 nm thick high-quality epitaxial Pb(Zr,Ti)O$_3$ (PZT) film was grown by Pulsed Laser Deposition (PLD) on a DyScO$_3$ (DSO) substrate. A 20 nm thick SrRuO$_3$ (SRO) layer was grown, before the PZT deposition, as the bottom electrode. 50 nm of Pt is then deposited and patterned using shadow masking technique to serve as the top electrode.
Figure 3f) depicts the schematic of the PZT capacitor together with its polarization characteristic.

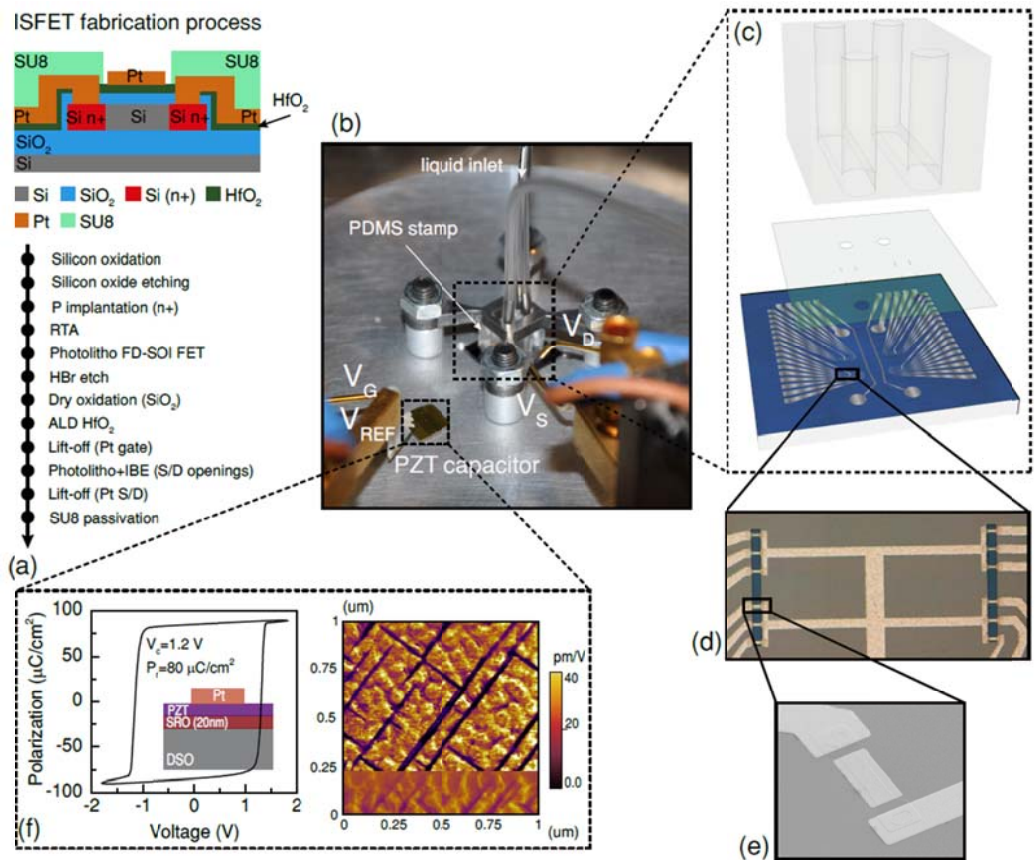

Figure 3 NC-ISFET experimental setup. a) Schematic process flow for the FD UTB SOI ISFET fabrication. b) Setup used for validating the NC-ISFET operation: the Gate bias is applied on one contact of the PZT-based Fe-Cap and the contact on the other side is externally connected to the integrated RE. c) Exploded 3D rendering of the system composed by the ISFET chip, the SU8 passivation layer and the PDMS stamp with microfluidics channels. d) Detail of a set of ISFET sensors with openings in the SU8passivation. e) SEM image of a single ISFET. f) Schematics, polarization characteristics and PFM image of the epitaxially grown PZT layer.

The polarization hysteresis loop measured at 100 Hz shows a remanent polarization of about 80 $\mu C/cm^2$ and an almost symmetric coercive voltage of 1.2 V. The Piezoelectric Force Microscopy (PFM) analysis confirms that the polarization in all c-domains is oriented from bottom to top interface. In order to test the responsivity of the ISFET to pH, a set of five pH buffers, ranging from pH 4 to pH 8, have been flown on the devices and for each of them the transfer characteristics has been plotted. A PDMS stamp and a foundry-made Aluminum support structure have been used to create an air-tight channel which allowed a controlled flow of the LUT on the sensing gates. The flow rate was set via an 11 Elite Harvard apparatus syringe pump. The Fe-Cap has been externally connected to the commercial flow-through RE (MI-16-701 Microelectrodes Inc.RE), as demonstrated in Figure 3b). After each measurement the channels have been emptied and DI water has been flown inside them using the same syringe pump employed for the pH buffers. After the DI water cleaning, the channels have been emptied again before flowing new buffers. In order to demonstrate that the performance improvements are due to the Negative Capacitance Effect, a set of tests has been performed on a Fe-Cap – ISFET configuration without capacitive matching as well (excessive area of the Fe-Cap).

**Results and discussion** The transfer characteristics and the corresponding extracted sensitivities of the conventional ISFET, Fe-ISFET and NC-ISFET are presented in Figure 4.

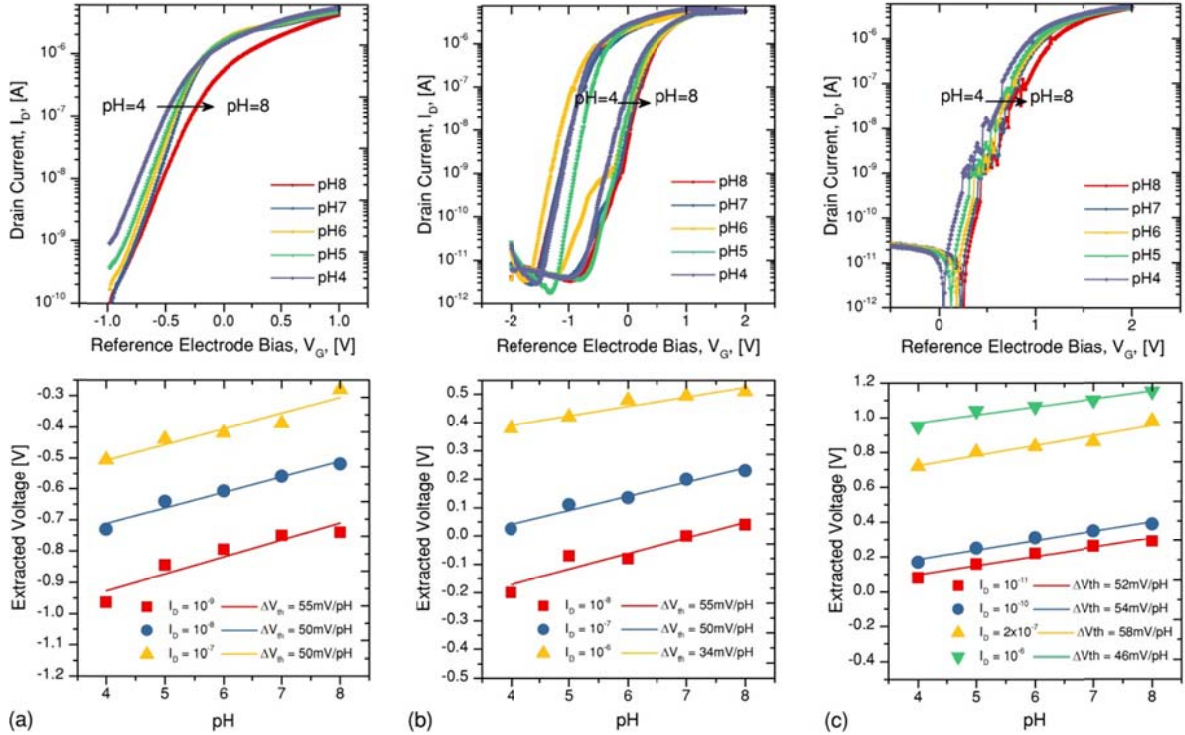

Figure 4 ISFETs characteristics. a-c) Transfer characteristics extracted at different pH (top) and pH sensitivities extracted at different ID levels (bottom) for (a) a conventional ISFET, (b) a Fe-ISFET and (c) a Nearly hysteresis-free capacitively matched NC-ISFET.

In case of Fe-ISFETs/NC-ISFETs (Figures 4b and 4c), the gate voltage is swept from -2 V to 2 V and back, while the drain voltage was set at 0.1 V. In case of the Fe-ISFET, when the condition for NC to occur is not fulfilled[29], the device shows a huge hysteresis without any boosting effect. Additionally, it should be remarked that a hysteretic device is not suitable for sensing applications as it provides a memory effect. On the other hand, in a device structure with a fully matched design of capacitances, where the total capacitance of the structure remains positive in the whole gate voltage range, hysteresis-free characteristics with boosting of both analog and digital performances of the ISFET is achieved (Figure 4c). The sensitivities of each sensor, in terms of mV per pH, have been extracted from the transfer characteristics at various current levels (Figure 4a-c, bottom). It can be noticed that this sensitivity, close to the ideal limit of 60 mV/pH is not degraded by the insertion of the Fe-Cap. It should be noted that no NC-related improvement was expected for this sensitivity, since the NC-effect, as stated, is acting on the second transduction mechanism alone (from Gate bias to Drain current), therefore the first one (from ion concentration to Gate bias) is subjected to the Nernst limit as normal. Figures 5a) and b) clearly demonstrate the improvement achieved in the SS of the ISFET: the average value is reduced from 196 to 110 mV/dec, showing an enhancement by 44%. This means that, for a pH variation of one point ($\Delta V_G$ = 55mV, according to the extracted sensitivities), the conventional

ISFET would have a relative Drain current increase of 1.9 times, while the same device with the addition of the NC element would show a Drain current increase close to 3.2 times.

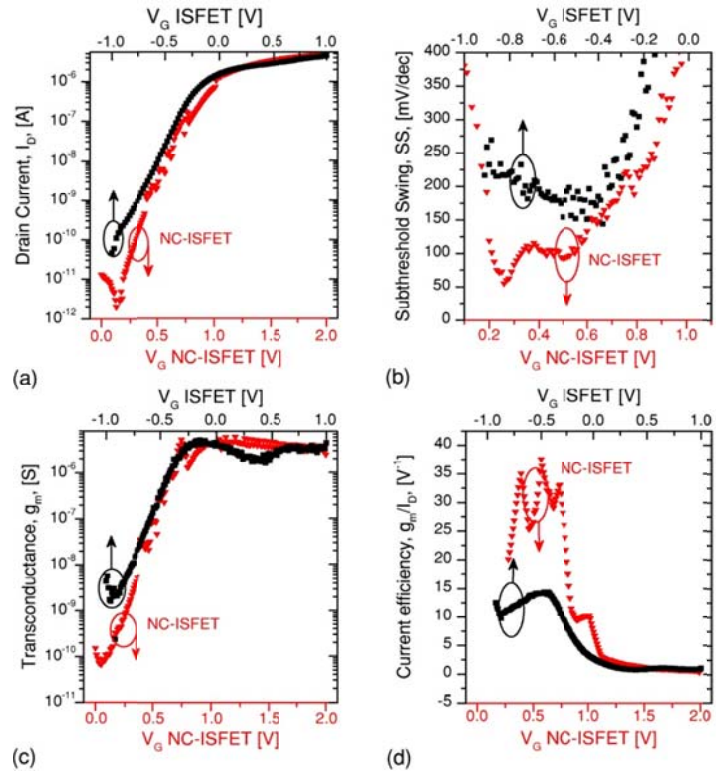

Figure 5 ISFETs comparison. a) Transfer characteristics comparison for conventional and NC-ISFET: the NC configuration brings to a steeper SS. b) SS comparison for conventional and NC-ISFET: in the weak inversion range, the average value for the NC configuration is 110mv/dec, while for the standard configuration it is close to 200. c) Transconductance comparison for conventional and NC-ISFET: the NC configuration shows a steeper variation. d) Current efficiency comparison for conventional and NC-ISFET: the NC configuration shows a value higher than twice that of the standard one.

Another relevant parameter to consider when evaluating the performances of a device, especially in the field of wearable or portable sensors, is the current efficiency $g_m/I_D$, corresponding to the efficiency with which charges are inserted in a system: the higher the carrier efficiency, the lower are the energy losses and therefore the power consumption. Figure 5d) shows that also in this field the insertion of the NC component brings relevant improvements, rising the average $g_m/I_D$ ratio by 2.5 times in the bias range relevant for sensing. The presence of the true NC effect in the sensor is confirmed by the differential gain $\Delta V_{int}/\Delta V_{out}$ on the Fe-Cap, shown in Figure 6.

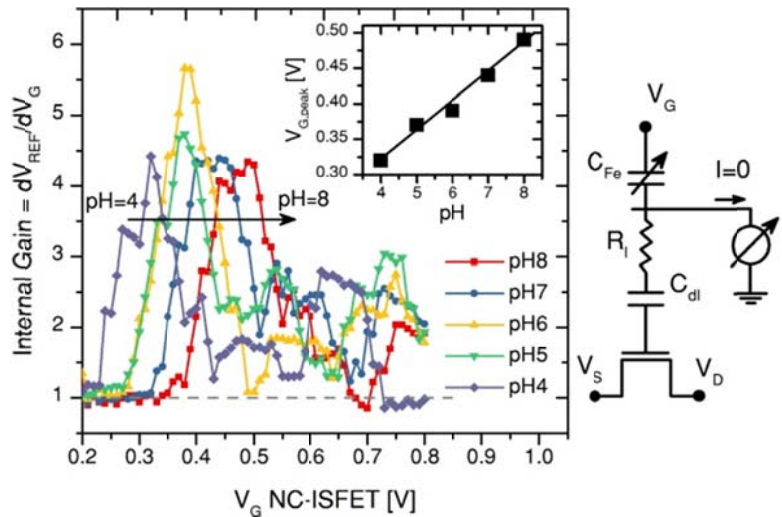

Figure 6 Gain-Voltage characteristics of the NC at different pH values. The peak magnitude is bigger than unity for all pHs, demonstrating the step-up effect in the bias applied to theRE. The position of the peak is dependent on the pH of the solution in contact with the RE and can therefore be employed as an alternative method to measure it (inset).

Negative Capacitance can be claimed if the ratio is bigger than unity: this is immediately verified for our system where, in agreement with our modeling results, the peak of the gain $\delta V_{REF}/\delta V_G$ lies between 4.5 and 5.5. Another interesting effect shown by this graph is that the peak of the gain is shifting with pH in the same way as the transfer characteristics, which makes it usable as a new metrics of sensitivity. The inset of Figure 6 demonstrates a linear dependence on pH, with a measured sensitivity of 41 mV/pH. Furthermore, the fact that the gain peak is shifting the same way as the transfer characteristics allows always having the maximum NC-boost in the relevant bias range. This effect can be understood from the Nernst equation for the dependence of the Ag-AgCl RE potential on the activity of the chloride ions[30]. It is important to underline that, since it is necessary to switch the Fe-Cap from one stable state to the other, NC is only observed when the applied bias is swept over a specific range of given width. In terms of device operation, this has two important consequences:
1. The NC-ISFET cannot be operated with a fixed bias (or, more precisely, can be operated only as a conventional ISFET), but rater needs a full sweep for each measurement.
2. Depositing the Ferroelectric material directly on the gate of the ISFET would not increase the $\Delta V_G$ for a given pH variation. For this reason, the NC-ISFET configuration proposed in this work is not only the easiest to implement, but also the most efficient (the same performance improvement is achieved with a single Fe-Cap instead of one per each sensor).

**Conclusions** In this paper we have proposed a new technique to apply the Negative Capacitance effect, which has already been proven as an effective performance booster of conventional MOS transistors, to improve sensitivity and current efficiency of an ISFET sensor. The measurements performed have shown that the inclusion of a Ferroelectric capacitor in series between the Gate bias source and the Reference Electrode has no detrimental effect on the ion sensing achieved by the gate material, but is able to both decrease the SS, improving the current sensitivity, and increasing the current efficiency, reducing the power consumption. A mathematical model for the capacitive matching with the electrolyte double-layer has been introduced and verified through a comparison with the actual NC gain measured in this work. We also showed that, when this capacitive matching is not achieved, the transfer characteristics show a strong hysteresis and no improvement of the SS.

The dependence of the gain peak position on the pH of the solution analyzed has been reported and explained in terms of the dependence of the RE potential on the activity of the chloride ions in the LUT. This secondary effect has been proposed as an additional method for the measurement of pH in a solution.

**References**

1. Bergveld, P. Development of an ion-sensitive solid-state device for neurophysiological measurements. *IEEE Transactions on Biomedical Engineering* **1**, 70–71 (1970).
2. Wipf, M. et al. Selective sodium sensing with gold-coated silicon nanowire field-effect transistors in a differential setup. *ACS nano* **7(7)**, 5978-5983 (2013).
3. Rigante, S. et al. Sensing with advanced computing technology: fin field-effect transistors with high-k gate stack on bulk silicon. *ACS nano* **9(5)**, 4872-4881 (2015).
4. Buitrago, E., Fagas, G., Badia, M. F. B., Georgiev, Y. M., Berthomé, M., & Ionescu, A. M. Junctionless silicon nanowire transistors for the tunable operation of a highly sensitive, low power sensor. *Sensors and Actuators B: Chemical* **183**, 1-10 (2013).
5. Buitrago, E. et al. Electrical characterization of high performance, liquid gated vertically stacked SiNW-based 3D FET biosensors. *Sensors and Actuators B: Chemical* **199**, 291-300 (2014).
6. Ahn, J. et al. A pH sensor with a double-gate silicon nanowire field-effect transistor. *Applied Physics Letters* **102(8)**, 083701 (2013).
7. Schwartz, M. et al. DNA detection with top–down fabricated silicon nanowire transistor arrays in linear operation regime. *Physica status solidi (a)* **213(6)**, 1510-1519 (2016).
8. Zafar, S., D'Emic, C., Afzali, A., Fletcher, B., Zhu, Y., & Ning, T. Optimization of pH sensing using silicon nanowire field effect transistors with HfO2 as the sensing surface. *Nanotechnology* **22(40)**, 405501 (2011).
9. Accastelli, E., Scarbolo, P., Ernst, T., Palestri, P., Selmi, L., & Guiducci, C. Multi-wire tri-gate silicon nanowires reaching milli-pH unit resolution in one micron square footprint. *Biosensors* **6(1)**, 9 (2011).
10. Chen, S., Bomer, J. G., Carlen, E. T., & van den Berg, A. Al2O3/silicon nanoISFET with near ideal Nernstian response. *Nano letters* **11(6)**, 2334-2341 (2011).
11. Kim, S. et al. Silicon nanowire ion sensitive field effect transistor with integrated Ag/AgCl electrode: pH sensing and noise characteristics. *Analyst* **136(23)**, 5012-5016 (2011).
12. Jang, H. J., & Cho, W. J. Performance enhancement of capacitive-coupling dual-gate ion-sensitive field-effect transistor in ultra-thin-body. *Scientific reports* **4**, 5284 (2014).
13. Bergveld, P. Thirty years of ISFETOLOGY: What happened in the past 30 years and what may happen in the next 30 years. *Sensors and Actuators B: Chemical* **88(1)**, 1-20 (2003).
14. Nakata, S., Arie, T., Akita, S., & Takei, K. Wearable, flexible, and multifunctional healthcare device with an ISFET chemical sensor for simultaneous sweat pH and skin temperature monitoring. *ACS sensors* **2(3)**, 443-448 (2017).

15. Douthwaite, M., Koutsos, E., Yates, D. C., Mitcheson, P. D., & Georgiou, P. A thermally powered ISFET array for on-body pH measurement. *IEEE transactions on biomedical circuits and systems* **11(6)**, 1324-1334 (2017).
16. Garcia-Cordero, E., Bellando, F., Zhang, J., Wildhaber, F., Longo, J., Guërin, H., & Ionescu, A. M. Three-Dimensional Integrated Ultra-Low Volume Passive Microfluidics With Ion Sensitive Field Effect Transistors For Multi-Parameter Wearable Sweat Analyzers. *ACS nano* (2018).
17. Angelidis, P. A. Personalised physical exercise regime for chronic patients through a wearable ICT platform. *International journal of electronic healthcare* **5(4)**, 355-370 (2010).
18. Ghafar-Zadeh, E. Wireless integrated biosensors for point-of-care diagnostic applications. *Sensors* **15(2)**, 3236-3261 (2010).
19. Lee, Y. H., Jang, M., Lee, M. Y., Kweon, O. Y., & Oh, J. H. Flexible field-effect transistor-type sensors based on conjugated molecules. *Chem* **3(5)**, 724-763 (2017).
20. Bruen, D., Delaney, C., Florea, L., & Diamond, D. Glucose sensing for diabetes monitoring: recent developments. *Sensors* **17(8)**, 1866 (2017).
21. Salahuddin, S. & Datta, S. Use of negative capacitance to provide voltage amplification for low power nanoscale devices. *Nano letters* **8**, 405–410 (2008).
22. Jo, J. et al. Negative capacitance in organic/ferroelectric capacitor to implement steep switching MOS devices. *Nano letters* **15**, 4553–4556 (2015).
23. Saeidi, A. et al. Negative capacitance as performance booster for Tunnel FETs and MOSFETs: an experimental study. *IEEE Electron Device Letters* **38(10)**, 1485-1488 (2017).
24. Appleby, D. J. et al. Experimental observation of negative capacitance in ferroelectrics at room temperature. *Nano letters* 14, 3864–3868 (2014).
25. Zubko, P. et al. Negative capacitance in multidomain ferroelectric superlattices. *Nature* **534**, 524–528 (2016).
26. Saeidi, A. et al. Negative capacitance field effect transistors; capacitance matching and non-hysteretic operation. In *Solid-State Device Research Conference (ESSDERC), IEEE* 2017, 78-81 (IEEE, 2017).
27. Velikonja, A., Gongadze, E., Kralj-Iglic, V., & Iglic, A. Charge dependent capacitance of Stern layer and capacitance of electrode/electrolyte interface. *Int. J. Electrochem. Sci* **9**, 5885-5894 (2014).
28. Starkov, A. & Starkov, I. Asymptotic description of the time and temperature hysteresis in the framework of Landau-Khalatnikov equation. *Ferroelectrics* **461**, 50–60 (2014).
29. Rusu, A., Saeidi, A., & Ionescu, A. M. Condition for the negative capacitance effect in metal–ferroelectric–insulator–semiconductor devices. *Nanotechnology* **27(11)**, 115201 (2016).
30. Preočanin, T., Šupljika, F., & Kallay, N. Evaluation of interfacial equilibrium constants from surface potential data: silver chloride aqueous interface. *Journal of colloid and interface science* **337(2)**, 501-507 (2009).

**Acknowledgements** The authors acknowledge funding of this research from Milli-tech/ERC H2020. The authors also greatly appreciate the contributions of Mr. Robin Nigon and Prof. Paul Muralt in the fabrication of the PZT thin film.